# A spectral approach to stock market performance

Escañuela Romana, Ignacio[1]

Escañuela Nieves, Clara[2]

**Abstract**

We pose the estimation and predictability of stock market performance. Three cases are taken: US, Japan, Germany, the monthly index of the value of realized investment in stocks, prices plus the value of dividend payments (OECD data). Once deflated and trend removed, harmonic analysis is applied. The series are taken with and without the periods with evidence of exogenous shocks. The series are erratic and the random walk hypothesis is reasonably falsified. The estimation reveals relevant hidden periodicities, which approximate stock value movements. From July 2008 onwards, it is successfully analyzed whether the subsequent fall in share value would have been predictable. Again, the data are irregular and scattered, but the sum of the first five harmonics in relevance anticipates the fall in stock market values that followed.

# 1. Introduction.

The study of the stock market, in different countries, with the aim of explaining and predicting the evolution of prices, volumes and dividends, occupies thousands of scientific articles with the application of very different perspectives and models. This interest is due, among other things, to the fact that financial markets have a strong impact on economies. For example, Basak et al. (2019) point out the importance of stock prices for the investment of firms that need external equity to finance their marginal investments. In addition, Barro and Ursúa (2009) emphasize the possibility of predicting economic depressions from stock market crashes. GARCH models, which use autoregression of shocks and lagged variances, are used to predict market volatility with relative success (Akgiray, 1989). Hence, expectations of returns are formed, based on the relationship between volatility and expected returns.

[1] Loyola University of Andalusia, Doctoral candidate iescanuelaromana@al.uloyola.es ignacioesro@gmail.com

[2] Max Planck Institute for Nuclear Physics, Heidelberg, Germany clara.escanuela@mpi-hd.mpg.de

Stock markets are an extraordinarily complex phenomenon that is very difficult to model, explain and predict because, as Basak et al. (2019) state, "It is well-known that stock market price series are generally dynamic, non-parametric, chaotic and noisy in nature making investments intrinsically risky" (p. 553). To give just one example, on 19 October 1987 the Dow Jones Industrial Average fell 508 points, a drop of 22.6% in one day (Shiller, 1987). Because of this complexity, the models developed explain some aspects, but leave many others unanswered and have limited empirical validity (Peters, 1996). The question of this paper is posed in Fama (1965): can we make estimates and predictions about the stock market price? It is a question of the utmost relevance and one that remains open.

Our approach to the answer is through harmonic analysis, applied to a stock market price index variable. We analyze three cases - the United States, Japan and Germany - after World War II, and test with the original data series, deflated by the CPI, and with the alternative of removing from it the periods in which we know of the existence of some exogenous shock. For all six series, then, a periodogram estimation is performed. Also, a prediction in which we place ourselves in July 2008 (the so-called Great Recession) and see whether the most relevant hidden periodicities, in terms of their contribution to the variance of the observed series, would have allowed us to predict the fall in stock market values that followed. We first carry out an empirical test of the random walk model, which denies the possibility of the predictability of stock market performance.

The scientific environment of this work, in which it acquires relevance, is formed by the discussion of the empirical validity of the random walk model and by the different alternative models to the previous one that have been proposed. In both cases, it is impossible in this article to carry out a sufficient minimum review of the empirical scientific literature and we limit ourselves to mentioning some of the most important works.

An initial consensus was established in the 1970s and 1980s around two hypotheses, set out, after numerous precedents (e.g., Howrey,1965, who studied the spectrum of stock price changes), in Samuelson (1965), Fama (1965, 1970) and Malkiel (2003). First, the efficient market hypothesis is stated for capital markets: a market is in an efficient position in relation to the available set of information. Therefore, an agent would not be able to predict the future evolution of the price of the shares exchanged in the market from past data, because this market would have already incorporated all the information in the formation of prices and the successive information of the future will be presented randomly (if there were a pattern, it would have already been incorporated). The "weak" version analyses whether agents could use past information to systematically improve their investment returns (Fama, 1970), admitting that some may be luckier than others but on average there is no such ability to improve prediction. This would be the consequence of multiple rational agents with access to all the information at a reduced cost (Fama, 1995).

Second, that movements in the prices of stocks would follow a random walk (RWM). The efficient markets hypothesis does not necessarily imply the random walk model, but it is by far the most widely used. It implies two basic ideas. First, that the successive price changes are independent, and the probability distribution of the price change in period t is independent of the sequence of prices in previous periods, so knowledge of these does not contribute to predicting the probability distribution for the price change during the next period (Fama, 1965):

$$\Pr(y_t = y | y_{t-1}, y_{t-2}, \dots) = \Pr(y_t = y) \qquad (1)$$

Where y is price change, and Pr is the probability that the price change takes y value. On the right-hand side of the equation, unconditional; on the left-hand side of the equation, conditional on knowledge of previous price changes. The equation is true as long as an agent cannot improve his prediction with knowledge of the past.

The second idea states that (Godfrey et al., 1964) random disturbances $\epsilon_t$, which are the difference between the values of the variables between t and t-1, follow different possible distributions, but always all of them have zero mean values. The distribution that satisfies the condition that the second moments are finite is discussed (Lo and MacKinlay, 1988).

The random walk equation is: $y_t - y_{t-1} = \epsilon_t \qquad (2)$

Where $y_t$ are prices at time t, $\epsilon_t$ is white noise, $E(\epsilon_t)=0$, $E(y_t \epsilon_t)=0$, and it is not autocorrelated with its past values. Therefore, he states that "the actions of the many competing participants should cause the actual price of a security to wander randomly about its intrinsic value" (Fama, 1995, p. 4). Intuitively, a large number of people are predicting price as best as possible, which is that prices in the future equal prices in the past (Granger and Morgernstern, 1963) and differences are random. The investor cannot beat the market in his predictions and any unanticipated patterns of behavior would then be incorporated by investors as a whole into the future, nullifying their effect.

Sometimes a drift, a weighted average of the probabilities of each future price occurring, is added to the right-hand side of the equation (Dupernex, 2007). If prices are expressed in logarithms, then we are in a geometric random walk model and the drift expresses a growth factor (Nau, 2014): constant required or expected return.

By the mid-1970s, these ideas seemed theoretically and empirically proven (Malkiel 2003; Dupernex, 2007). Kendall (1953) studied 22 series of industrial share prices and concluded that in the vast majority the random changes are much more pronounced than any systematic movement that might exist. Fama (1965, 1970) and Malkiel (2003) argue that the market does not follow a mathematically perfect REM and independence will never be complete, but that the observed dependence is very small, so much so that no investor can use knowledge of the past to establish an improved investment plan, compared to an investor investing randomly. Biases would cancel out and only RWM is founded on a coherent and unifying theory.

However, numerous empirical studies since the 1990s have been divided in their results. Lim and Brooks (2011) point out in their survey that these studies have focused on the predictability of security returns on the basis of past price changes, or the profitability of trading strategies based on past returns. They point to the relevance of the school of behavioral finance, which would introduce a number of cognitive and behavioral biases that lead real stock markets away from RWM. Lo (2005), Adam and Merkel (2019), among others, cite frequent behaviors such as risk aversion, overconfidence, over-reaction, etc. which could lead to speculative mispricing situations. Lo and MacKinlay (1988) and Chen (1996) find that the price series of securities quoted in the market would be autocorrelated. Granger and Morgenstern (1963) point out that stocks follow a random walk in the short run, but in the long run they have non-stochastic components. Okpara (2010), Ajekwe et al. (2017) studied the Nigerian stock market and concluded that this stock market was weak form efficient and followed a random walk process. However, Dittrich and Srbek (2020) analysed the daily returns of selected stock indices and individual firms and concluded that the hypothesis of a random walk in stock prices is not correct. Dias et al. (2020) analyse several stock exchanges and their performance in the Covid19 pandemic, such as the US, China, Germany and others, and conclude that there is partial evidence against the efficient market hypothesis. Lo (2004) proposes an adaptive approach where biases in investment behavior or departures from the efficient markets hypothesis are explained as evolutionary approximations over time to perceived circumstances. In conclusion, as Leković (2018) points out, the debate on the empirical validity of the efficient markets hypothesis is not over.

On the other hand, different alternative models have been proposed to predict stock market performance. Christoffersen and Diebold (2003) reaffirm the possibility of making predictions on which to base investment strategies. Caporin et al. (2013) conclude that high and low prices of equity are predictable (through a vector autoregressive model). In the real world of investing, strategies such as the fundamental approach (earnings ratios, company strategies, leverage ratios, etc.) and the technical approach or graphical and correlation analysis (Thakkar and Chaudhari, 2021). Lo, Mamaysky and Wang (2000) find that "technical" analysis, also called "charting", appears to have practical value in increasing stock market investment returns. In addition, economic research has proposed multiple approaches, in two broad groups. One is the behavioral finance perspective, which proposes the use at the time of investment of predictions based on heuristic patterns used by agents, e.g. Thaler (1999) and Shiller (2000). The latter mentions investor confidence and the existence of bubble expectations and finds certain changes in this perception over time. The second would be studies using computational intelligent methods and algorithms, as detailed in the survey by Cavalcante et al. (2016). Fischer and Krauss (2018) apply sequential learning, through the long short-term memory networks technique, while Thakkar and Chaudhari (2021) evaluate the results of deep neural network techniques. Ashley and Patterson (1989) use a bispectral test to detect a non-linear dynamic mechanism that would generate the stock market index.

Special mention should be made of the mean reversion model and its central idea that the stock price series may move away from the mean or trend value (due to economic or financial movements), but eventually return to that value (due to movements in company decisions). The basic equation (Ornstein-Uhlenbeck process) would be as follows (Metcalf and Hassett, 1995, p. 4):

$$dP_t = \propto (\bar{P} - P_t)P_t dt + \sigma P_t dz \qquad (3)$$

$P_t$ prices at time t, $\bar{P}$ is the mean, dz is shock to prices, α is a coefficient measuring the speed of reversion. It is therefore claimed that the series reverts to its trend, at a rate proportional to the distance of the values away from its trend, and that it is therefore stationary in the long run. This paper argues that, indeed, the stock price series does not follow a random motion, but adds, with respect to this mean reversion model, a stronger claim: that the series follows regular cyclical movements that harmonic analysis can find.

Our perspective is based on harmonic analysis. A number of papers apply it to the stock market. Among others, as mentioned above, Granger and Morgenstern (1963) conducted a spectral analysis of New York stock market prices, concluding that the short-term frequencies roughly follow a RWM, but the longer-term, longer-period components are more relevant than that model predicts. However, in the results of the spectral analysis, Granger (1966) considered that its use cannot give interesting results if the economic time series measured in level have spectra that exhibit a smooth declining shape with considerable power at very low frequencies. Then the results obtained should not have such characteristics. Uebele and Ritschl (2009) use multivariate spectral analysis on the movements of national output, nominal wage, business activity and the stock market before World War I in Germany to fix the chronology of the business cycle. Escañuela (2011) applied harmonic analysis to the Dow Jones Industrial Stock Average series (United States). Chaudhuri and Lo (2015) use the cross-periodogram to analyze, in the U.S. stock market, the volatility and correlation of U.S. common-stock returns, while Syukur and Marjuni (2019) apply spectral analysis with a Hadamard transform for Stock Price Forecasting in the 30-day time horizon and Shu and Zhu (2019) apply the Lomb spectral analysis on the detrended residuals of two Chinese stock indices. Finally, articles such as Beaudry et al. (2020) use spectral analysis to study aggregate economic series, in the case of this article for data series on labour intensity, such as the unemployment rate, hours worked per capita.

**2. Data.**

(a) United States, Japan and Germany. Share price index. From OECD (2022a), 2015 base year. 1957:01 to 2022:07 (United States), 1959:01 to 2022:07 (Japan), 1960:01 to 2022:07 (Germany). Calculated from the prices of common shares of companies traded on national or foreign stock exchanges (determined by the stock exchange, using the closing daily values for the monthly data). The index adds on to the price index the value of dividend payments, assuming they are re-invested in the same stocks. We chose the index for its overall value in terms of the country's companies and the incorporation of dividends to get the effective value of the investment. However, its performance is very close to that of other stock market indices. These series are henceforth referred to as the "stock market" of each country. We deflate the indices using the consumer price index of each country, base 1960:01.

(b) United States, Japan and Germany. Consumer price index. From OECD (2022b), 2015 base year. Defined as the change in the prices of a basket of goods and services that are typically purchased by specific groups of households.

On the other hand, we find a number of exogenous shocks that we consider to be external to the stock market but have an impact on it:

i. United States. 1962:10 Cuban missile crisis. Year 1974, 1979:11 to 1980:06 and 1999:12 to 2003:03 an increase in oil prices of more than 75% in the previous twelve months, 2001:09 to 2002:01 terrorist attacks in the United States, 2003:02 to 2003:04 Iraq war. Covid pandemic from 2020:02.

ii. Japan, Germany. Same periods of shocks. Except: in both we consider that the shock due to the terrorist attacks ends two months earlier than in the US, in Germany we consider external shock to the process towards reunification, 1990:03 to 1990:12, in Japan Fukushima nuclear accident 2011:03 to 2011:04.

The selection of the periods with exogenous disturbances, the factor considered to determine them and the onset and duration of these disturbances are subject to ad hoc selection criticisms. Why one period and not another? It is a matter of taking and eliminating those moments in which non-economic causes could have an impact on the observed series. In any case, the results obtained are approximately the same whether or not these disturbances are eliminated.

## 3. Method.

Harmonic analysis is applied to time-stationary series, signal statistics like mean and variance do not change over time (as opposed to alternatives such as multifractal spectrum). Non-stationary series are generally made stationary by removing trend and seasonality. The drawback of this procedure is that the trend may be a cycle of longer period than the length of the series being removed. It is possible to assume the existence of a trend in the mean and to eliminate it (Morgenstern, 1961). We understand the existence of a trend conceptually as defended by Álvarez Vázquez (1986): "The wave hypothesis states that it is simply another cyclic component, although through insufficient knowledge, that is, limitation in the size of the series, we do not know it in its totality" (p. 7). In addition, there is debate about the form of the trend. The frequently used forms are first differences, the string method and least squares. We chose to fit the best polynomial least squares equation and remove its values from the series. The reason for the selection is to take the mean of the series that best fits the data, even at the risk of, as Álvarez Vázquez (1986) points out, eliminating part of the information of the cycle. The method used in this paper has many points in common with that used in Escañuela and Escañuela (2022).

We use two fundamental instruments to analyze the behavior of the series studied. First, the autocorrelation function (ACF). Second, the Fourier analysis to extract hidden periodicities of the series. The ACF is the correlation between one time series $x_t$ and a lagged one $x_{t+k}$. The definition of autocorrelation for lag k is:

$$\rho_k = \frac{Cov(x_t, x_{t-k})}{\sqrt{Var(x_t)Var(x_{t-k})}} \quad (4)$$

Where the numerator is the covariance between two values of the series separated by a lag k at time t. The denominator is the normalization of the covariance with variances. When the series is stationary, the denominator is simply the time-independent variance of the series ($\sigma^2$).

The object of the spectral analysis is to search for periodicities in stationary series. The frequency of the wave ($f$) is defined as the inverse of the period ($T$) in time units. Likewise, the angular frequency $\omega$ is $2\pi f$ in radians. A stationary time series may be described as a trigonometric function,

$$x_t = \sum_{j=1}^{n/2} A_j \cos(2\pi f_j t + \varphi_j) \qquad (5)$$

$A_j$ is a coefficient and $\varphi_j$ is a phase factor. The time series have been modelled with this alternative and also equivalent equation, $x_t = \sum_{j=1}^{n/2} \left( a_j \cos 2\pi \frac{j}{n} t + b_j \sin 2\pi \frac{j}{n} t \right) \qquad (6)$

$f_j = \frac{j}{n}$ are the Fourier or harmonic frequencies, a and b are coefficients. The harmonic frequencies lie between 0 and 0.5 inverse of time units. Corresponding to values for $\omega$ between 0 and $\pi$. The aim is to find the harmonic frequencies with the highest contribution to the signal modelling. The Schuster periodogram is based on Fourier transformations which converts a signal in time domain to one in the frequency domain. Discrete Fourier tranforms (DFT) are defined as follows:

$$d(f_j) = \frac{1}{\sqrt{n}} \sum_{t=1}^{n} x_t e^{-2\pi f_j t} \qquad (7)$$

Which is computationally performed with the Fast Fourier Transform (FFT) method with Python (language used to perform the calculations and graphs in this work). The periodogram at a frequency $f_j$ is,

$$I(f_j) = |d(f_j)|^2 = \frac{n}{4} A_j^2 \qquad (8)$$

The scaled periodogram $P(f_j)$ is simply the periodogram multiplied by $4/n$. It is interesting how the spectral density can be derived from the Fourier transform of the autocorrelation. Another important relationship is given by Parseval's theorem (Chatfield, 2013): $\frac{1}{n} \sum_{j=1}^{n} (x_t - \bar{x})^2 = \sum_{j=1}^{n/2-1} \frac{A_j^2}{2} + a_{n/2}^2 \qquad (9)$

On the left the variance of the series and $\frac{A_j^2}{2}$ is the contribution to the variance by the jth harmonic frequency. This expression is really important since it implies that the power spectral density measures directly the contribution of each harmonic frequency to the variance of the series. The periodogram, then, allows one to estimate the percentage of the variance of the analyzed data series as a function of each of the discrete frequencies. From this result it is concluded that the data would not be periodic if the contribution to the variance is negligible. But it is periodic if it is necessary to consider sine and cosine waves and one or more harmonic frequencies to explain the data points and make accurate predictions (Moran, 1950).

The periodogram is asymptotically unbiased (as the series grows, the periodogram tends to its true value), but not consistent (the variance of $I(f_j)$ does not tend to zero). A consistent estimator can be achieved by averaging over close frequencies which implies loss of information. Problems with the periodogram are, for example, noise and resolution. Other periodograms have been developed to overcome the limitations of the traditional periodogram.

The Welch periodogram method consists of dividing the data series into smaller segments that may or may not overlap with each other. The periodogram of each of these segments is calculated and then averaged. In this method, as in many others, a window is used. A window is a function that multiplies the series to account for the finiteness of the data. Welch's method decreases noise, losing ability to resolve frequencies. As we alter the window, from rectangular to Hanning (as an example), we see a higher signal-to-noise ratio and worse resolution (Jwo, Chang and Wu, 2021, p. 3992).

On the other hand, the Lomb-Scargle periodogram is a much used method is astronomy as it allows one to model data points which are irregularly distributed over time. It can be applied when, for example, some data points are missing due to external phenomena. It is, furthermore, situated at the intersection point between the four ways through which current periodograms are defined. It is given by Fourier analysis and the least squares method. But it can also be derived from the principles of Bayesian probability and can sometimes be related to bin-based phase-folding techniques (VanderPlas, 2017).

In addition, we apply tests that are important for the purpose of this article. The goodness of fit of the model and prediction is measured with these tests. Firstly, the Pearson's chi-squared test which is calculated as follows,

$$\chi^2 = \sum_{t=1}^{n} \frac{(Y_t - E_t)^2}{E_t} \qquad (10)$$

Where Y is the measured value and E is the expected value. Models and predictions are also tested with the so-called symmetric Mean Absolute Percentage Error (sMAPE),

$$sMAPE = \frac{1}{n} \sum_{t=1}^{n} \frac{|Y_t - E_t|}{(|Y_t| + |E_t|)} \qquad (11)$$

This corrects the MAPE test but it is still not completely symmetric and it slightly overweights overpredictions. Other two standard accuracy tests are the Mean Absolute Error (MAE) and the Root Mean Square Error (RMSE):

$$MAE = \frac{1}{n} \sum_{t=1}^{n} |Y_t - E_t| \qquad (12)$$

$$RMSE = \sqrt{\frac{1}{n} \sum_{t=1}^{n} (Y_t - E_t)^2} \qquad (13)$$

## 4. Results.

We perform an empirical test of the RWM. In order for the entire series of, for example, the United States to be described with a RWM, the first differences must represent a white noise process. Fig. 1 shows the first differences of the series and compares them with simulated white noise. Visually there are clear differences. Therefore, we take the first differences and evaluate the autocorrelation (Fig. 2) and periodogram (Fig. 3) of the resulting series. Several delays are significantly non-zero and the periodogram is not fully uniform as expected for a white noise series. The Box-Pierce and The Ljung-Box tests conclude that the series is not purely white noise. The periodogram of the stationary (detrended) simulated RWM is shown in Fig. 4. We wish to compare this spectral density with the periodogram of the real detrended data.

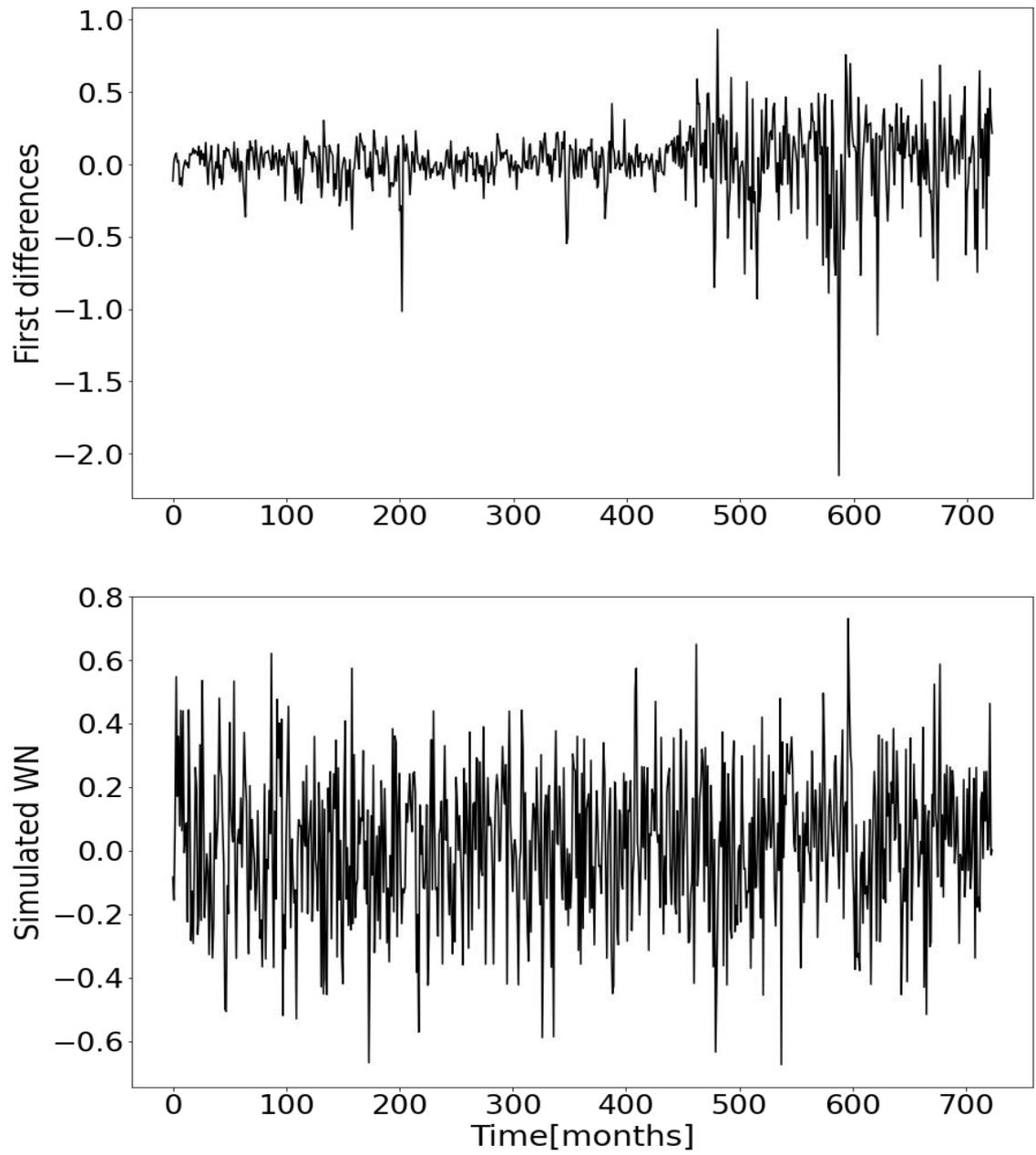

Fig. 1. First differences of the series (up) and simulated white noise (down).

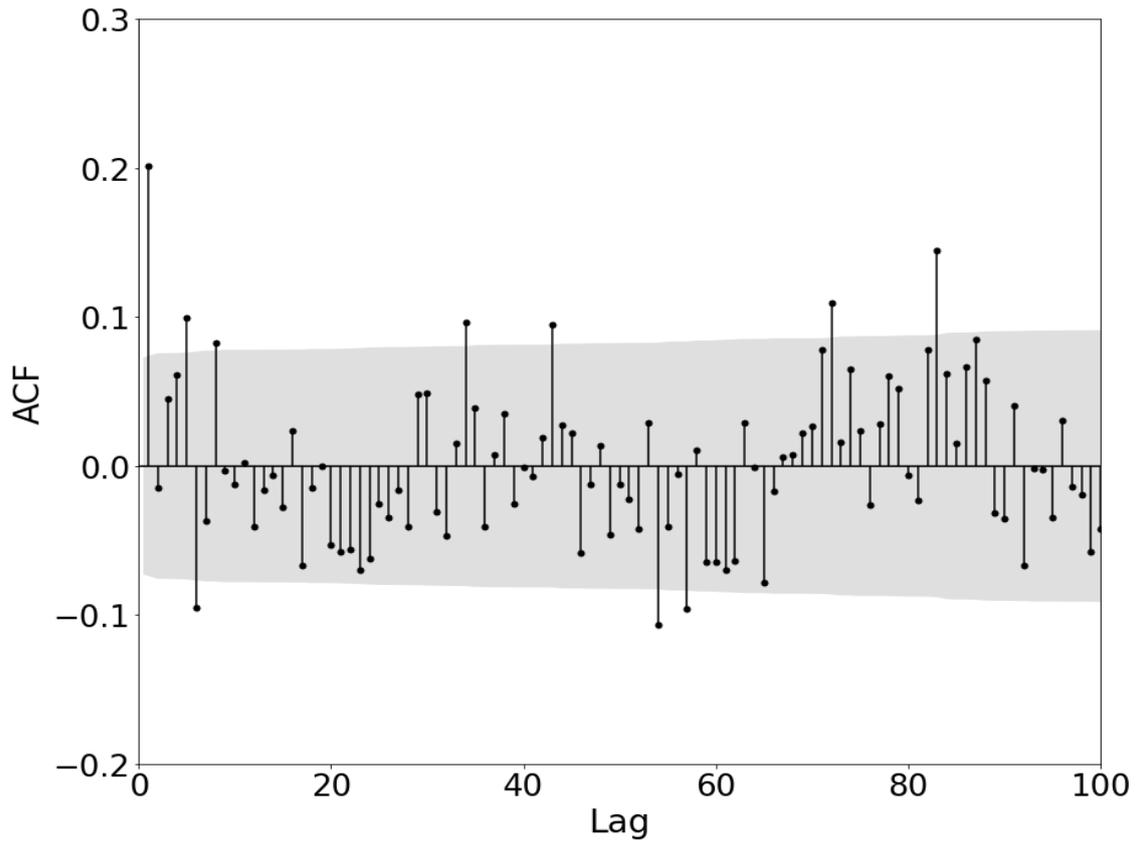

Fig. 2. Correlogram of the first differences of the time series of US.

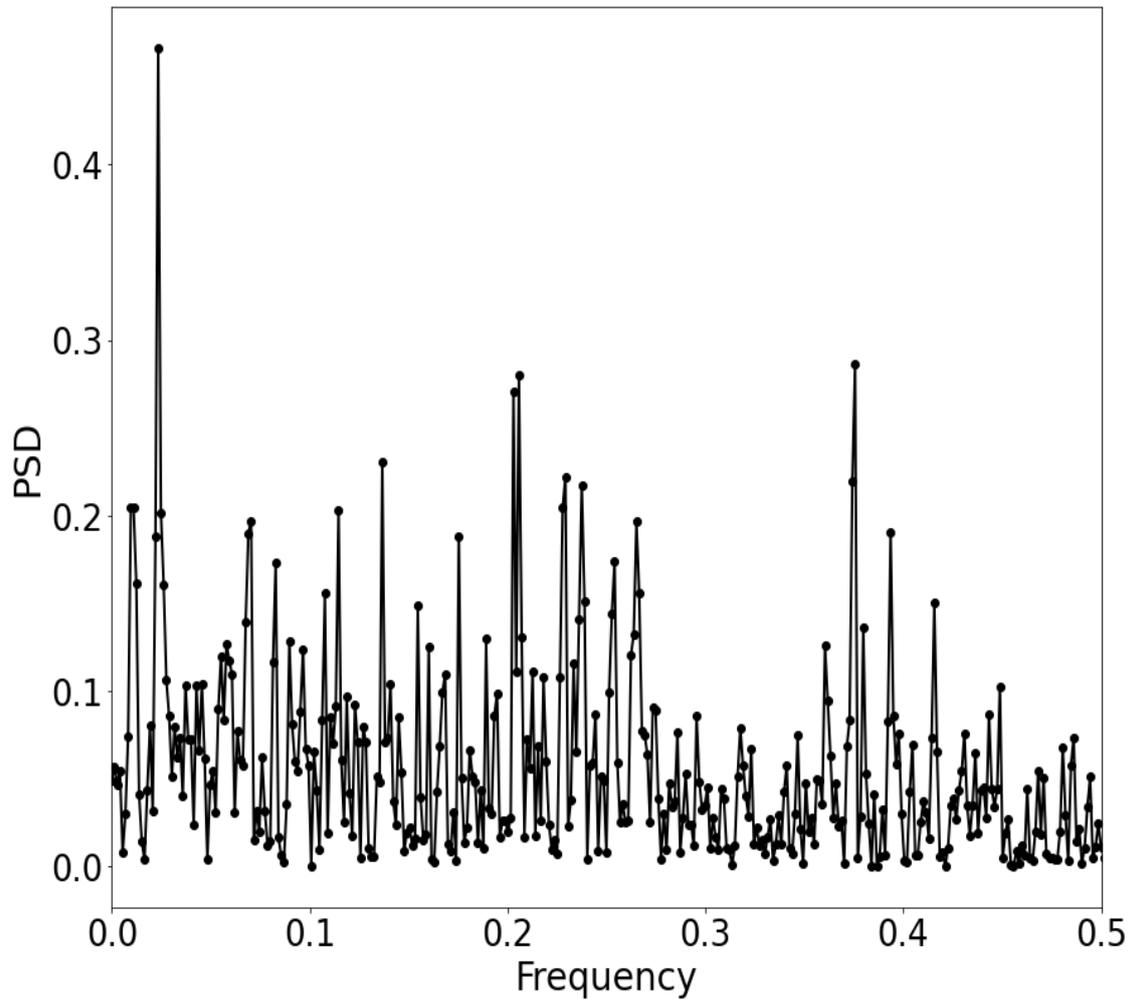

Fig. 3. Periodogram of the first differences of US.

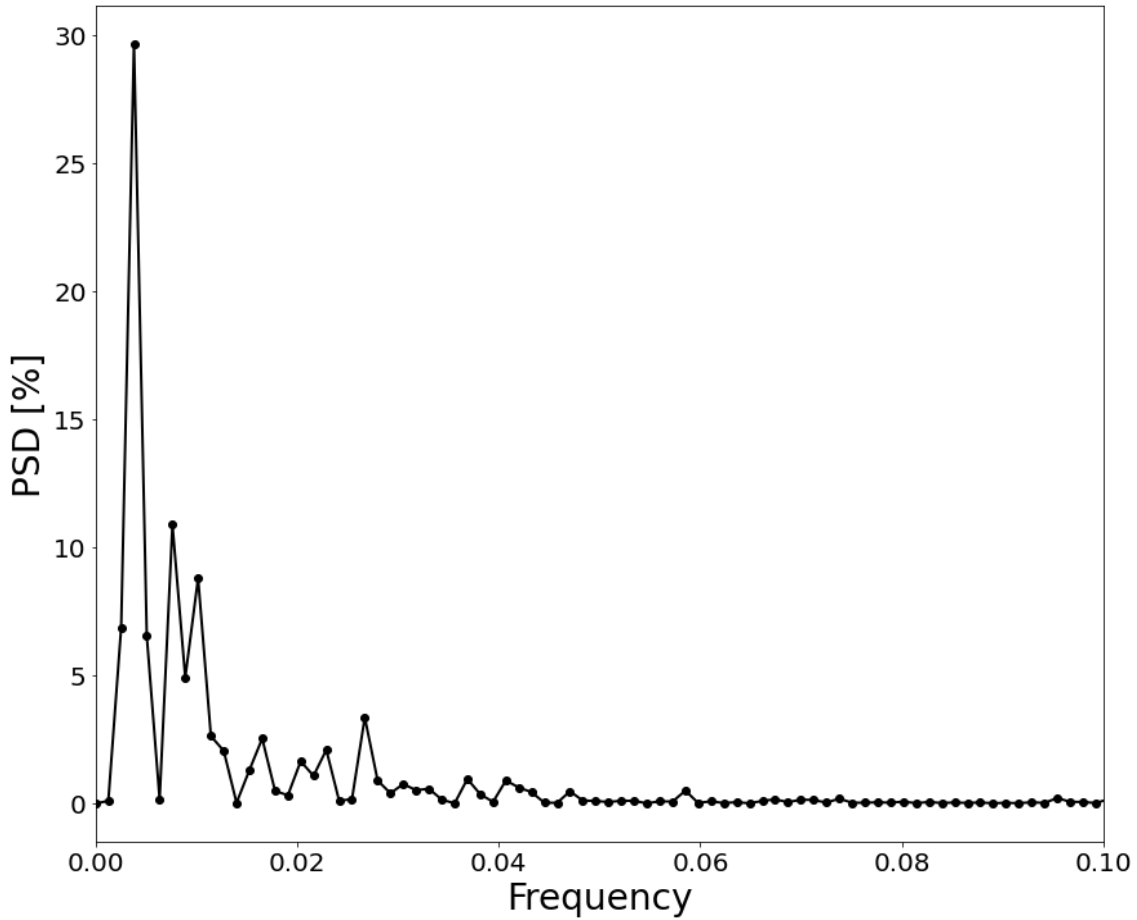

Fig. 4. Periodogram of detrended simulated random walk series ($y_t$).

In addition, there is a variable relationship between the three exchanges considered (Fig. 5). The German and US series are very similar: a very similar trend, coincident peaks and troughs. A high correlation of 0.95. Japan, however, shows a clearly different behavior, although it becomes more similar to the US as time progresses. A correlation of 0.43. The similarity between the cycles of the series can also be seen in the Welch periodogram of each series (Fig. 6), which have roughly the same main frequency peaks and similar contributions to the variance (y-axis), with Japan having the most cycles.

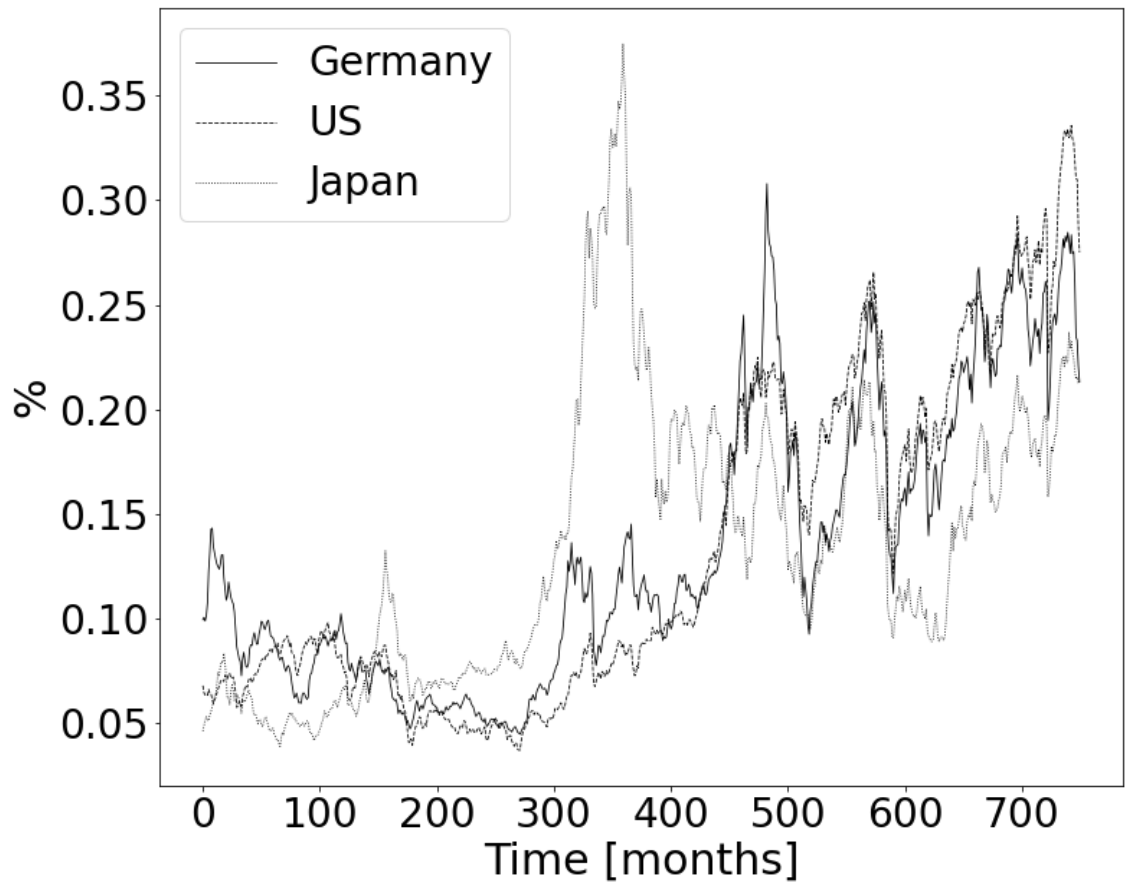

Fig. 5. Data in percentage of the US, Germany, and Japan stock exchange.

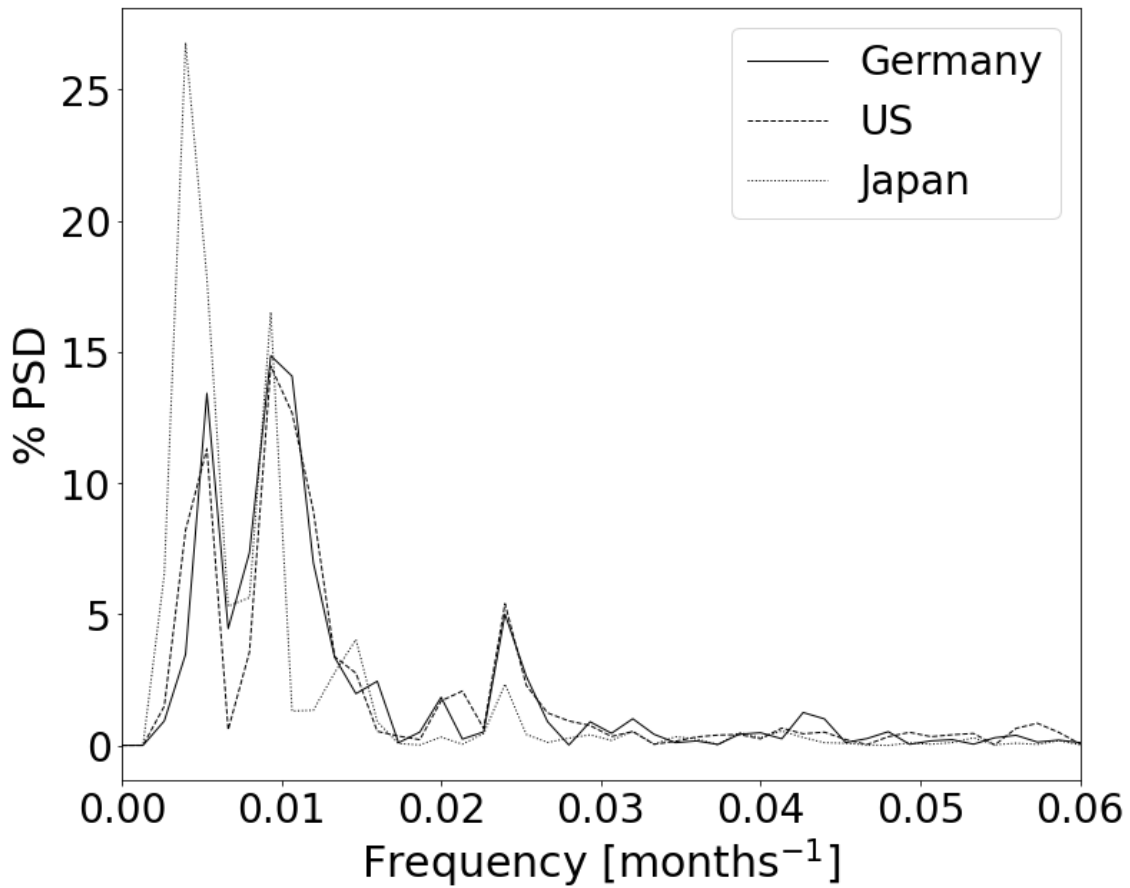

Fig. 6. Welch's method of peridogram for the three series between January 1960 and May 2022.

      Prior to the calculation of the periodogram, we have removed the trend: the US series is detrended with a fifth order polynomic equation (Fig. 7) to achieve a stationary series. Similarly, the data from Germany is detrended with the same order of polynomic equation that fits the trend better than for US, which could suggest that as the length of the series increases the trend deviates from a polynomic equation. It is logical to think that the trend is not really a polynomial as otherwise we are saying that stock market values increase indefinitely. Japan follows a different trend: sixth order polynomic equation. The definition of trend for each series is limited by the length of the series. It is almost inevitable that it will change as more data is collected. One possibility is the presence of long period cycles that have not been observed given the length of the observed series (data sparsity).

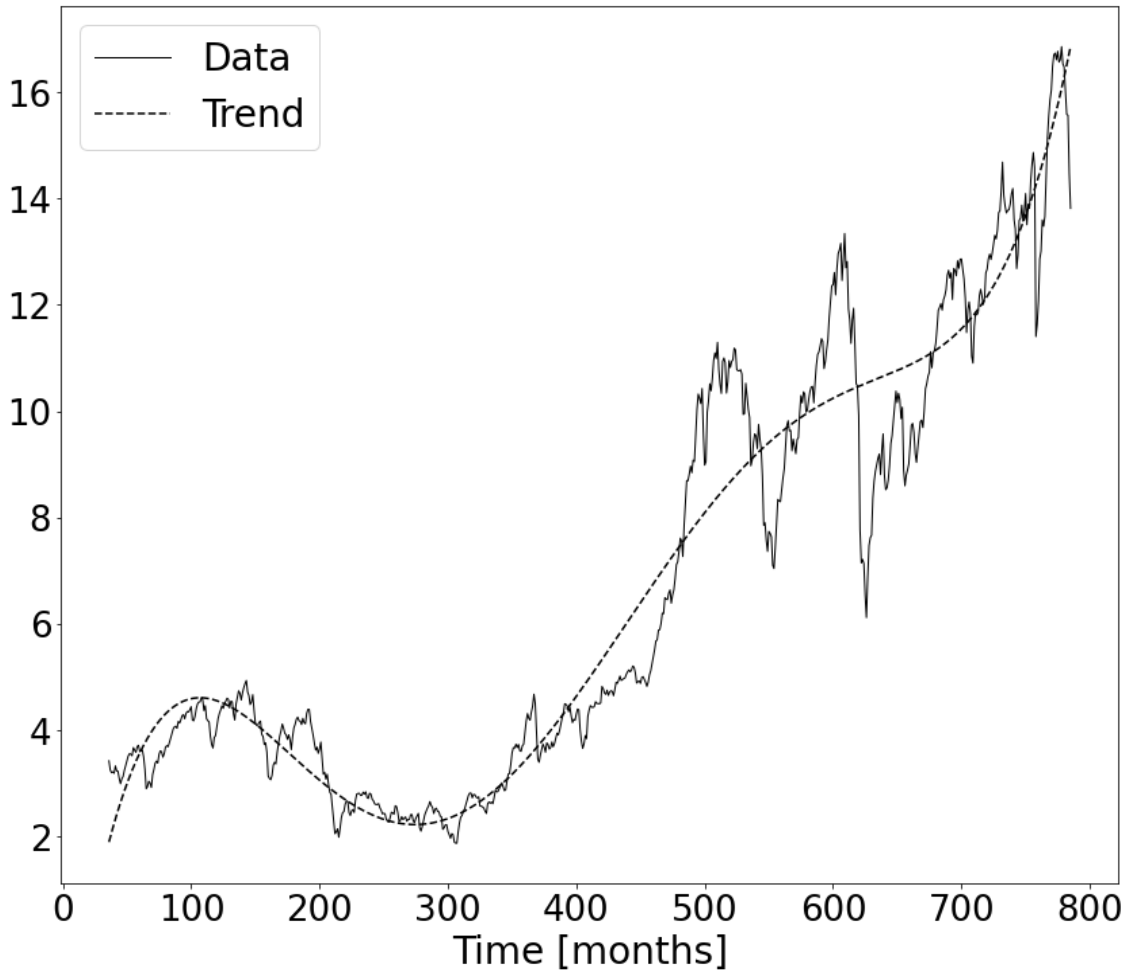

Fig. 7. US original time series with its trend (blue) by using least squares method.

The results of the periodogram with the Welch method and the Hanning window are as follows. The models for US, Germany and Japan are represented in Fig. 8, 9 and 10 respectively. The five most contributing frequencies are considered. In the last two years, atypical behavior is observed in the US and Germany due to the covid-19 crisis. This does not seem to have affected Japan as much.

These models are repeated with the Lomb-Scargle periodogram. The accuracy of these models is found in Table 1. The predictions are shown in Fig. 12, 13 and 14 and their accuracy, according to the tests and measurements presented in the results, in Table 2.

Table 1. Model accuracy.

|  | US | Germany | Japan |
| --- | --- | --- | --- |
| Pearson chi squared | Welch: -720 (dof*: 771)<br>Lomb: -300 (dof: 708) | Welch: 572 (dof: 735)<br>Lomb: 418 (dof: 661) | Welch: -642 (dof: 746)<br>Lomb: -2000 (dof: 681) |
| sMAPE | Welch: 50<br>Lomb: 46 | Welch: 50<br>Lomb: 45%. | Welch: 42%<br>Welch: 42%<br>Welch: 42%<br>Welch: 42%<br>Welch: 42%<br>Welch: 42<br>Lomb: 42%. |
| MAE | Welch: 0.49<br>Lomb: 0.47 | Welch: 1.35<br>Lomb: 1.37 | Welch: 1.52<br>Lomb: 1.77 |
| RMSE | Welch: 0.72<br>Lomb: 0.65 | Welch: 1.82<br>Lomb: 1.80 | Welch: 1.94<br>Lomb: 2.27 |
| Residual normality | Both: Not normal** | Both: Not normal | Both: Not normal |

*dof: degrees of freedom
**Normality checked with Kolmogorov-Smirnov (KS) test and Jarque-Bera (JB) test

Table 2. Prediction accuracy. With Lomb-Scargle periodogram.

| | US | Germany | Japan |
|---|---|---|---|
| Pearson chi squared | -261 (dof: 123) | 341 (cf: 123) | -312 (dof: 121) |
| sMAPE | 57% | 78% | 59% |
| MAE | 0.72 | 2.58 | 1.85 |
| RMSE | 0.97 | 3.32 | 2.47 |
| Residual normality | Normal (KS)<br>Not normal (JB) | Not normal (KS)<br>Normal (JB) | Not normal (KS)<br>Normal (JB) |

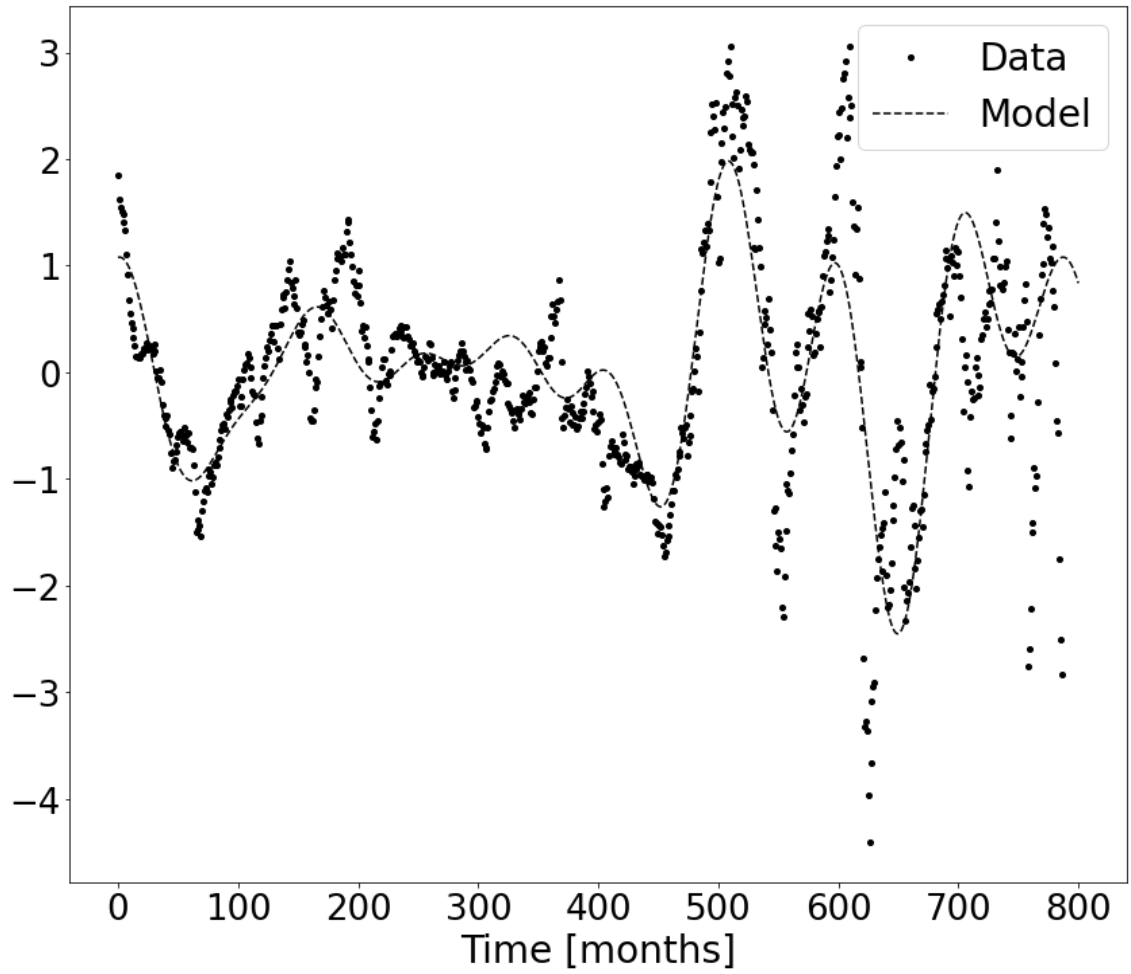

Fig. 8. US model with Welch's periodogram. Values between January 1957 and July 2022. The five most contributing periods are: 87.3, 112.3, 262.0, 196.5 (13%), and 98.25 (maximum contribution to the variance of 15%) months.

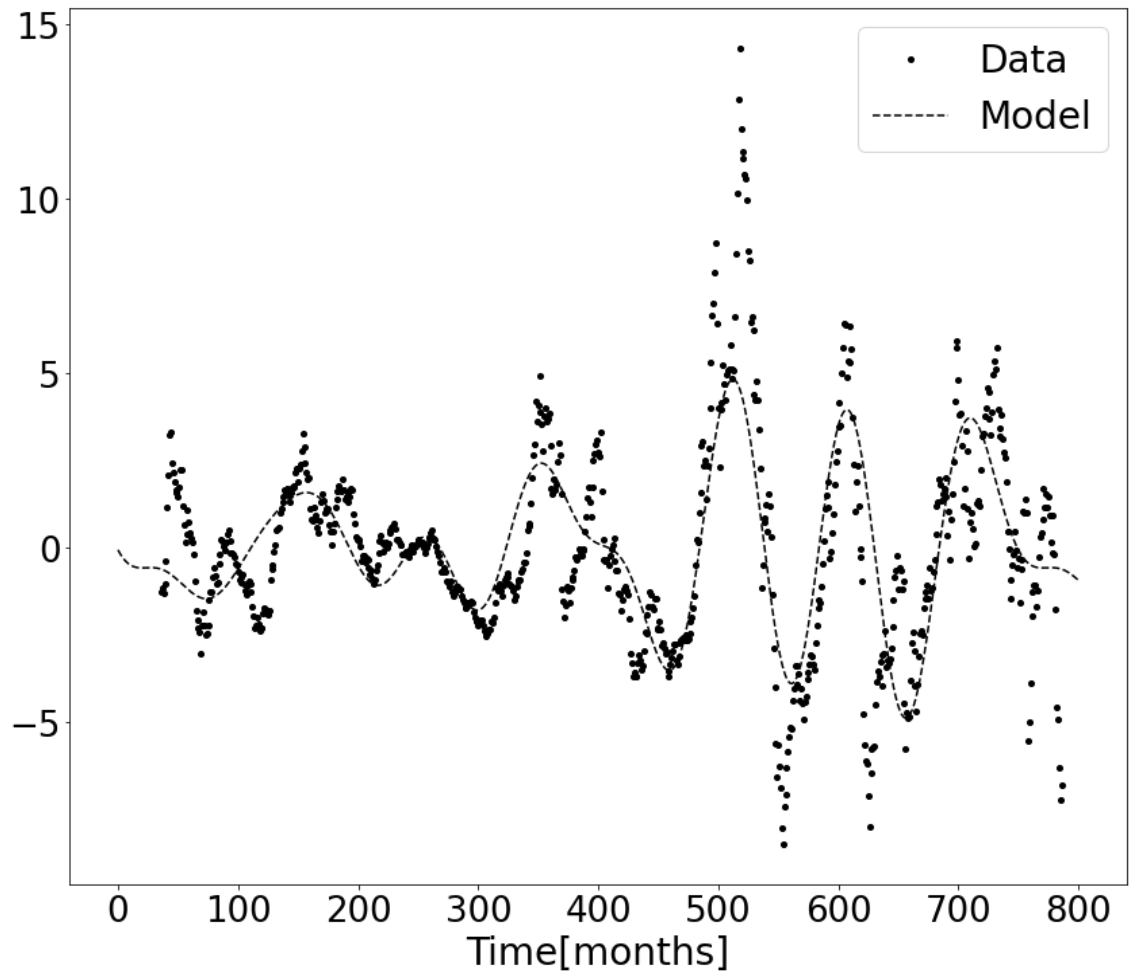

Fig. 9. Germany with Welch's method. January 1960 to July 2022. Periods: 83.3, 125.0, 187.5, 93.8 (14%), 107.1 (maximum contribution to the variance of 15%) months.

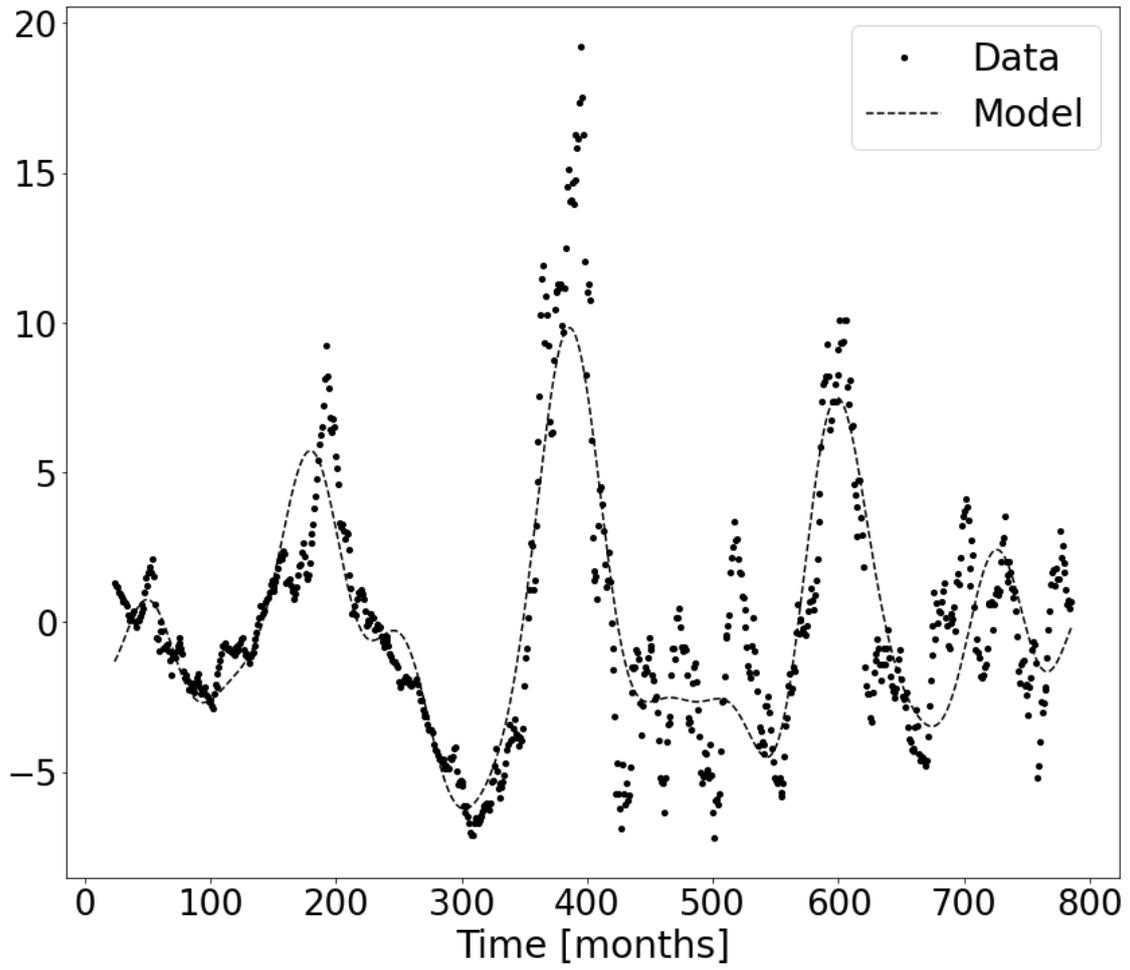

Fig. 10. Japan with Welch's method. January 1959 to July 2022. Periods: 125.0, 68.2, 250.0, 107.1 (18%), 187.5 (33%) months.

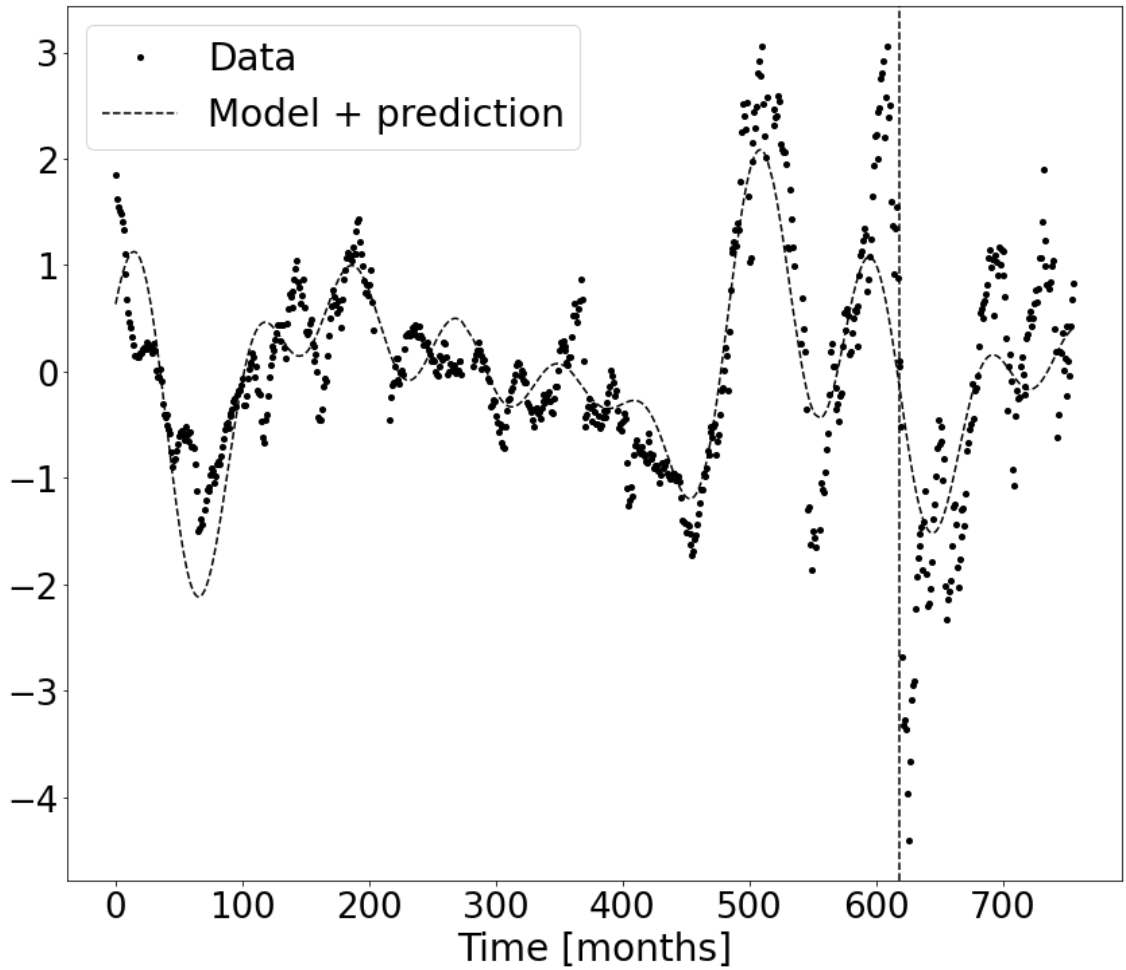

Fig. 11. Prediction from July 2008. Data with no exogeneous shocks. US. Periods: 42.2, 83.1, 116.3, 98.2, 184.4 (with maximum contribution of 16% to the variance) months. Dotted line: July 2008.

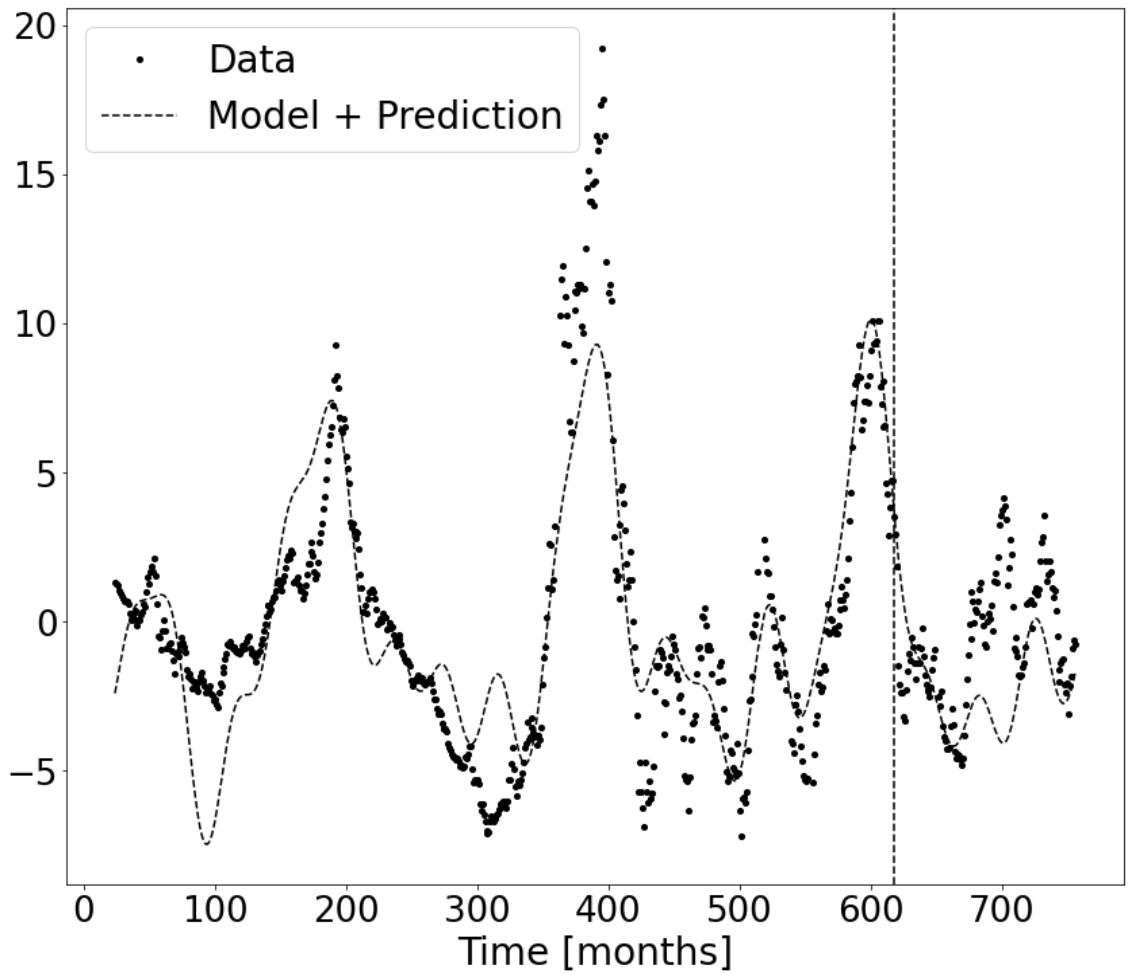

Fig. 12. Japan. Prediction from July 2008. No exogeneous shocks. Periods: 40.8, 78.0, 67.9, 107.1, 203.9 months. Maximum at 54% variance contribution (That is for period 203.9). Prediction seems worse at the beginning of the series and improves. It is able to predict the crisis.

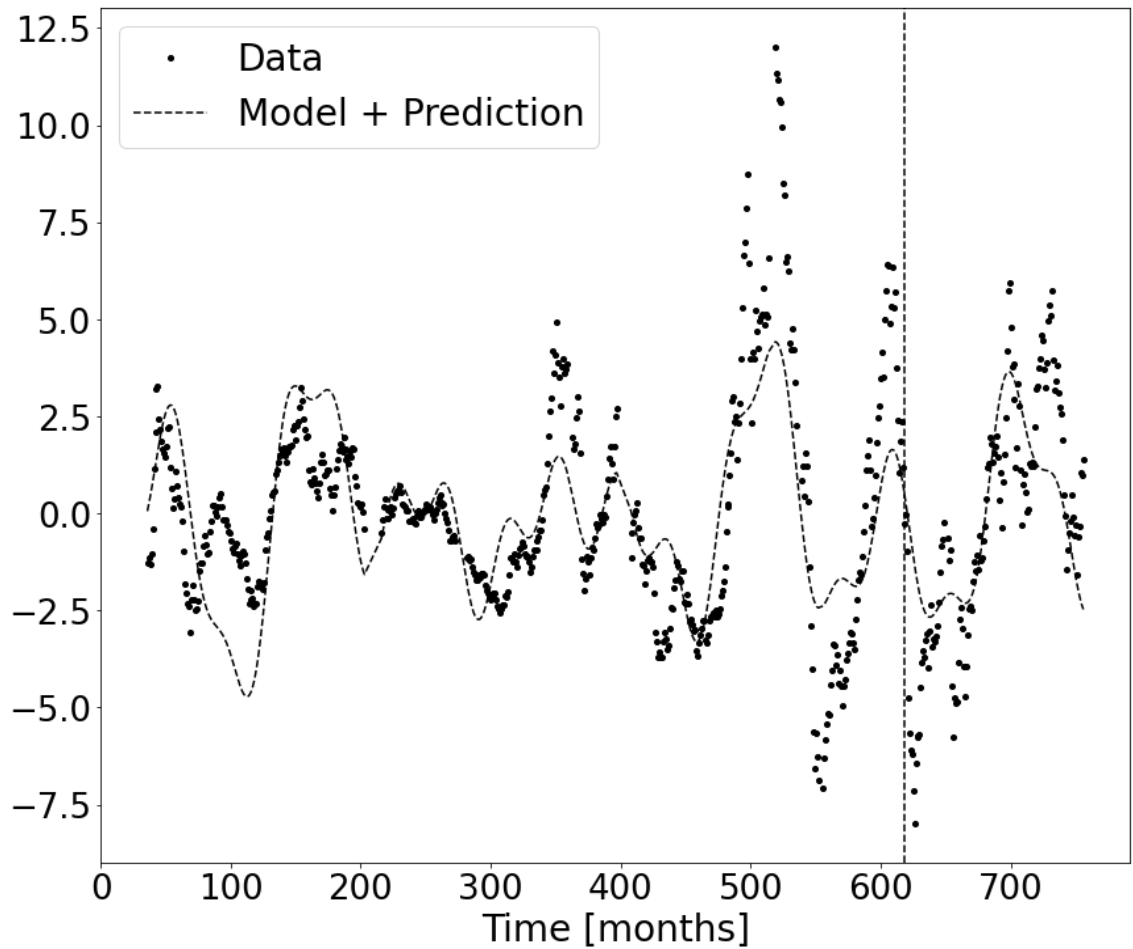

Fig. 13. Germany. Prediction from July 2008. No exogeneous shocks. Periods: 287.8, 42.7, 169.7, 112.2, 90.68493151 (Maximum with 20% contribution) months.

**5. Discussion.**

First, autocorrelation and harmonic analysis indicate that stock markets do not behave like a random walk, as there is relevant serial correlation in the series of first differences and the spectrum is not flat.

Second, the harmonic analysis shows the existence of relevant hidden periodicities. This provides the basis for an analysis of regular cyclical movements that is much more assertive than Mean Reversion type models (equation 3). The five most important harmonics make a significant contribution to the variance of the observed series, after extraction of a trend calculated by regression of a polynomial equation. The US and German stock markets seem to share four of the five main harmonics: approximately 7, 10, 16 and 8 years. The Japanese stock market seems to share both the 10- and 16-year periodicities. It is noteworthy that if the US and German stock markets move mainly together, the Japanese stock market shows a different behavior.

Third, the fit of the estimate is far from perfect as indicated by the tests and shown graphically. However, the sum of the five largest harmonics generally indicates the trend of the exchanges analyzed.

Fourth, the prediction in the most difficult case, that of the 2008 crisis, shows a fit that, again, is far from perfect. However, it is noteworthy that in all three cases it correctly predicts the downward movement of stock prices. In other words, the hidden periodicities do not perfectly adjust the future evolution of stock prices in the market, but they do capture the general movement of stocks: they correctly predict the subsequent market crisis.

**6. Conclusion.**

This paper rejects the validity of the application of the RWM to the case of stock price series (and embedded dividends) on the US, German and Japanese stock exchanges. It also finds the validity of a harmonic analysis approach applied to these series, which does not allow us to estimate each of the movements recorded in these series, since they are highly dispersed and irregular, but does allow us to estimate the most important movements.

This is confirmed by the 2008 crisis prediction experiment. While the sum of the main harmonics is not able to predict every single stock market price position, which appears to overreact, it correctly anticipates a bear market.

We understand that the interpretation of the relevance of low frequencies in the periodogram, which allows us to estimate and predict fundamental movements, can be put in the terms of Godfrey et al. (1964): in such long periods investors do not calculate profits, as the period exceeds their investment horizon, and this means that prices do not evolve according to a random walk. This would lead to a lagged response of the market prices to the underlying factors governing those prices (Alexander, 1961).